\input epsf
\documentstyle[12pt,a4]{article}
\begin{document}
\title {Unusual light spectra from a two-level atom in squeezed vacuum}
\author {G.C. Hegerfeldt, T.I. Sachse, and D.G. Sondermann\\\it
         Institut f\"ur Theoretische Physik, Universit\"at G\"ottingen\\
         D-37073 G\"ottingen, Germany}
\date {July 7, 1997}
\maketitle
\begin{abstract}
We investigate the interaction of an atom with a multi-channel squeezed vacuum.
It turns out that the light coming out in a particular channel can have
anomalous spectral properties, among them asymmetry of the spectrum,
absence of the central peak as well as central hole burning
for particular parameters.
As an example plane-wave squeezing is considered.
In this case the above phenomena can occur for the light spectra
in certain directions.
In the total spectrum these phenomena are washed out.
\vskip\smallskipamount\noindent PACS numbers: 42.50.Dv, 32.80.-t
\end{abstract}
\section{Introduction}
The interaction of squeezed light with atoms has found considerable
theoretical attention in the last years.
As pointed out in the seminal paper of Gardiner \cite{gasq}, squeezed light,
either in the white noise limit or broadband colored noise \cite{gaco},
can drastically alter the radiative properties of atoms and can,
in principle, reduce substantially the spectral linewidth of emitted
light.
Gardiner's paper has stimulated a large amount of work on the interaction
of squeezed light with atoms.
We refer to the detailed review of Parkins \cite{park} for references
and an account of work up to 1993.
For more recent work see, e.g.,
Refs.~\cite {p,fismysw}
and references therein.

Gardiner \cite{gasq,gabu} considered a one-dimensional like situation
in which only a single channel is squeezed
and where the atom couples only to this channel.
Here a channel means a set of modes ranging over all frequencies with the
remaining quantum numbers characterizing the modes kept fixed \cite{gasq}.
For example, a channel may consist of all modes with fixed angular momentum
and parity for a multipole expansion of the field, or of all plane wave
modes with a given direction and polarization.

An investigation of resonance fluorescence spectra of atoms interacting with
a (single-channel) squeezed vacuum plus a laser
was initiated by Carmichael, Lane, and Walls \cite{clw}
(for further references see the review \cite{park} and
Refs.~\cite {p,fismysw}).
Smart and Swain \cite{smsw,swain} have pointed out the existence of
interesting structures in these spectra.
In Ref.~\cite {buzek,fd} multi-channel squeezing and associated correlation
functions for three-level atoms were studied.

In this paper we consider the spectral effects of a multi-channel
squeezed vacuum in the white noise limit on a two-level atom.
For the atomic correlation functions and the {\em total}
spectrum of all outcoming light a multi-channel squeezed vacuum
leads to analogous results as a single-channel squeezed vacuum
with appropriate parameters.
However, in a multi-channel situation one can observe not only
the total spectrum but also the light spectrum in
individual channels.
It turns out that --- due to interference of the (quantized)
light scattered from the atom with the squeezed vacuum ---
these spectra can show unexpected features which are
not visible in the total spectrum, e.g.,
a possible asymmetry, absence of the central peak as well as central
hole burning for particular parameters.
By the same arguments as in Ref.~\cite {gaco} we expect these features
to persist also for only approximate white-noise squeezing.

In Section~\ref{II} below the general case of multi-channel squeezed
white noise interacting with a two-level atom is treated.
The spectrum is calculated in terms of a background, scattered, and
interference part.

In Section~\ref{III} we treat in detail an example in which the channels
consist of plane waves with fixed directions and polarizations.
In this case one has a divergence of the background term
when calculated in terms of photon numbers,
and we therefore use the spectral Poynting vector to calculate
the spectrum for given position and direction of observation of
the spectral analyzer.
In this case the above phenomena like asymmetry, central hole burning
etc.\ can occur for the light spectra in certain directions.

In Section~\ref{IV} we discuss our results in detail, in particular the
question of interference, and point out a possible
connection, probably more formal than
directly physical, with the results of Ref.~\cite {swain}.

\section{Spectra for general multi-channel squeezing}
\label{II}
We consider a two-level atom coupled to the electromagnetic field in
three-dimensional space.
The free Hamiltonians of the atom and the field are given by
\begin {equation}
H_{\rm A}=\hbar\omega _0\,\sigma ^+\sigma ^- \qquad
H_{\rm F}=\int_0^\infty{\rm d}\omega \sum_\alpha
  \hbar\omega \,a_\alpha ^\dagger (\omega )a_\alpha (\omega ) \;,
\end {equation}
where $\omega _0$ denotes the atomic transition frequency,
$\sigma ^+=|+\rangle\langle-|$ and $\sigma ^-=|-\rangle\langle+|$
are the atomic raising and lowering operators
and the $a_\alpha (\omega )$ are the annihilation operators of the field
obeying
the commutation relations
\begin {equation}
\label {acom}
[a_{\alpha _1}(\omega _1),a_{\alpha _2}^\dagger (\omega _2)]=
  \delta_{\alpha _1\alpha _2}\,\delta(\omega _1-\omega _2) \;.
\end {equation}
The index $\alpha $ stands for all quantum numbers of the chosen modes apart
from their frequency and thus characterizes a channel.
These are, for example, parity and angular momentum quantum numbers
if the multipole expansion of the field is used, or the direction of
propagation together with the polarization for plane wave modes.
In the rotating-wave approximation, the interaction Hamiltonian
may be written as
\begin {equation}
\label {haf}
H_{\rm AF}={{\rm i}\hbar\over\sqrt{2\pi}}\int_0^\infty{\rm d}\omega \sum_\alpha
  g_\alpha (\omega )\,a_\alpha ^\dagger (\omega )\sigma ^-\;+\;\hbox{h.c.}
\end {equation}
with possibly complex coupling coefficients $g_\alpha (\omega )$, which we
decompose as
\begin {equation}
g_\alpha (\omega )\equiv\sqrt{\gamma _\alpha (\omega )}\;{\rm e}^{{\rm
i}\phi_\alpha (\omega )} \;.
\end {equation}

Going over to the interaction picture leads to the Hamiltonian
\begin {equation}
H_{\rm I}(t)={\rm i}\hbar\sum_\alpha \sqrt {\gamma _\alpha }\,
\left[b_\alpha ^\dagger (t)\sigma ^--b_\alpha (t)\sigma ^+\right]
\end {equation}
with $\gamma _\alpha \equiv\gamma _\alpha (\omega _0)$ and
\begin {equation}
\label {bsa}
b_\alpha (t)\equiv{1\over\sqrt{2\pi}}\int_0^\infty{\rm d}\omega
  \left({\gamma _\alpha (\omega )\over\gamma _\alpha (\omega
_0)}\right)^{\!1/2}
  {\rm e}^{-{\rm i}\phi_\alpha (\omega )}\,
  {\rm e}^{-{\rm i}(\omega -\omega _0)t}\,a_\alpha (\omega ) \;.
\end {equation}
Assuming that the requirements for applying the Markov approximation
are satisfied we replace the second factor in the commutator
\begin {equation}
\label {bcom}
[b_\alpha (s),b_\beta^\dagger (t)]=
\delta_{\alpha \beta}\cdot
  {1\over2\pi}\int_0^\infty{\rm d}\omega \,{\gamma _\alpha (\omega )\over\gamma
_\alpha (\omega _0)}\,
  {\rm e}^{-{\rm i}(\omega -\omega _0)(s-t)}
\approx\delta_{\alpha \beta}\cdot \delta(s-t)
\end {equation}
by a $\delta$-function (cf.~\cite{ga,gabu}).

The radiation field is supposed to be initially in a pure broadband
squeezed vacuum state with the atomic frequency $\omega _0$ as central
frequency.
In general, for such a state the second order moments of the
$b_\alpha (\omega )$ read in the white noise limit \cite{ga,gabu}
\begin {equation}
\label {som}
\langle \,b_\alpha (s)b_\beta(t)\,\rangle ={\cal M}_{\alpha \beta}\,\delta(s-t)
\qquad
\langle \,b_\alpha ^\dagger (s)b_\beta(t)\,\rangle ={\cal N}_{\alpha
\beta}\,\delta(s-t)
\end {equation}
with
\begin {equation}
\label {emmenn}
{\cal M}^T={\cal M}\qquad {\cal N}^\dagger ={\cal N}\qquad {\cal M}^\dagger
{\cal M}={\cal N}({\cal N}+1) \;.
\end {equation}
If one omits factors of the form $[\gamma _\alpha (\omega )/\gamma _\alpha
(\omega _0)]^{1/2}$
this leads to
\begin {eqnarray}
\langle \,a_\alpha (\omega _1)a_\beta(\omega _2)\,\rangle &=&{\cal M}_{\alpha
\beta}\,
  \delta(2\omega _0-\omega _1-\omega _2)\,
  {\rm e}^{{\rm i}[\phi_\alpha (\omega _1)+\phi_\beta(\omega _2)]} \nonumber \\
\langle \,a_\alpha ^\dagger (\omega _1)a_\beta(\omega _2)\,\rangle &=&{\cal
N}_{\alpha \beta}\,
  \delta(\omega _1-\omega _2)\,
  {\rm e}^{{\rm i}[-\phi_\alpha (\omega _1)+\phi_\beta(\omega _2)]}
\end {eqnarray}
for the moments of the $a_\alpha (\omega )$.
In the following Eq.~(\ref {som}) will be used as definition
of squeezed white noise.

We shall assume in the following that ${\cal M}$ and ${\cal N}$ are diagonal
for the given modes,
\begin {equation}
\label {3dim}
{\cal M}_{\alpha \beta}=\delta_{\alpha \beta}\,M_\alpha \qquad
{\cal N}_{\alpha \beta}=\delta_{\alpha \beta}\,N_\alpha \;.
\end {equation}
One can imagine a state of this kind as being produced by independently
squeezing modes with different $\alpha $, e.g., by coupling them to
different parametric oscillators%
\ \cite{indep}.
For a pure state, as considered here, the assumption (\ref {3dim})
seems not to be very restrictive.
See \cite{spn,spnpre}, where for a finite number of modes the question
is discussed to what extent second order moments can be simplified by
choosing appropriate modes.

The fluorescence spectrum of an atom illuminated by squeezed white
noise has been studied by Gardiner \cite{gasq} and has since
become textbook material \cite{gabu}.
In Gardiner's article and in a large part of the following work it was
supposed that initially only a single channel of the radiation field is
squeezed and that the atom couples only to this channel.
In our notation, this means there is a particular $\alpha $, $\alpha =0$, say,
with
\begin {equation}
\label {1dim}
M_\alpha =M\delta_{\alpha ,0} \qquad
N_\alpha =N\delta_{\alpha ,0} \qquad
\gamma _\alpha =\gamma \delta_{\alpha ,0} \;.
\end {equation}
Such a situation will, a little imprecisely, be called
{\it one-dimensional} in the following.

We now want to calculate the spectrum ${\cal S}_\alpha (\omega )$ that
would be observed by a spectral analyzer coupled to the modes
$(\alpha ,\omega )$ with $\alpha $ fixed.
(For plane wave modes this would simply be the spectrum of light
with a certain polarization observed in a certain direction.)
One could of course also observe and determine other spectra, e.g.,
the spectrum in a channel which is a superposition of different $\alpha $'s.
Our procedure easily carries over to this situation.
To determine ${\cal S}_\alpha (\omega )$ we shall adapt the procedure of
Gardiner
\cite{gasq,gabu} to multi-channel squeezing.

At a finite time the spectrum is proportional to the expectation value
of the photon number operator $a_\alpha ^\dagger (\omega )a_\alpha (\omega )$.
The stationary spectrum is given by
\begin {equation}
\label {sft12}
{\cal S}_\alpha (\omega )=\lim_{T\to\infty}{1\over2\pi T}
  \int_0^T{\rm d}t_1\int_0^T{\rm d}t_2\,{\rm e}^{-{\rm i}(\omega -\omega
_0)(t_1-t_2)}
  \,w_\alpha (t_1,t_2) \;,
\end {equation}
with the two-time correlation function
\begin {equation}
w_\alpha (t_1,t_2)=\langle \,b_\alpha ^\dagger (t_1)_{\rm out}\,b_\alpha
(t_2)_{\rm out}\,\rangle \;,
\end {equation}
and can be written as the Fourier transform
\begin {equation}
\label {sft}
{\cal S}_\alpha (\omega )=
{1\over2\pi}\int_{-\infty}^\infty{\rm d}\tau \,{\rm e}^{-{\rm i}(\omega -\omega
_0)\tau }w_\alpha (\tau )=
{1\over\pi}\mathop {\rm Re}\int_0^\infty{\rm d}\tau \,{\rm e}^{-{\rm i}(\omega
-\omega _0)\tau }w_\alpha (\tau )
\end {equation}
of the stationary correlation function
\begin {equation}
w_\alpha (\tau )=\lim_{t_2\to\infty}w_\alpha (\tau +t_2,t_2) \;.
\end {equation}
The operators $b_\alpha (s)_{\rm out}$ appearing here denote the limits
\begin {equation}
b_\alpha (s)_{\rm out}=\lim_{t\to\infty}b_\alpha (s)_{\textstyle t} \;,
\end {equation}
where the subscript $t$ stands for the time evolution
in the interaction picture,
\begin {equation}
X_{\textstyle t}=U_{\rm I}(t,0)^\dagger \,X\,U_{\rm I}(t,0) \;.
\end {equation}
Since the $b_\alpha (s)_{\textstyle t}$ obey the equations of motion
\begin {equation}
{{\rm d}\over{\rm d}t}\,b_\alpha (s)_{\textstyle t}=
{{\rm i}\over\hbar}\left[H_{\rm I}(t)_{\textstyle t},b_\alpha (s)_{\textstyle
t}\right]=
\delta(t-s)\sqrt {\gamma _\alpha }\,\sigma ^-_{\textstyle t}
\end {equation}
one has
\begin {equation}
\label {bsat}
b_\alpha (s)_{\textstyle t}=b_\alpha (s)+\vartheta (t-s)\sqrt {\gamma _\alpha
}\,\sigma ^-_{\textstyle s} \;,
\end {equation}
where $\vartheta $ denotes the Heaviside function, and in particular
\begin {equation}
\label {out}
b_\alpha (s)_{\rm out}=b_\alpha (s)+\sqrt {\gamma _\alpha }\,\sigma
^-_{\textstyle s} \;.
\end {equation}
Inserting this into Eq.~(\ref {sft12}) yields a decomposition of the spectrum
into three parts, corresponding to the correlation functions
\begin {displaymath}
w_\alpha ^{\rm B}(t_1,t_2)\equiv\langle \,b_\alpha ^\dagger (t_1) b_\alpha
(t_2)\,\rangle \qquad
w_\alpha ^{\rm S}(t_1,t_2)\equiv\gamma _\alpha \langle \,\sigma ^+_{\textstyle
t_1}\sigma ^-_{\textstyle t_2}\,\rangle \qquad
\end {displaymath}\begin {equation}
\label {bis}
w_\alpha ^{\rm I}(t_1,t_2)\equiv\sqrt {\gamma _\alpha }\,\langle \,b_\alpha
^\dagger (t_1)\sigma ^-_{\textstyle t_2}+
\sigma ^+_{\textstyle t_1}b_\alpha (t_2)\,\rangle \;, \\
\end {equation}
which will be called the background, scattered, and interference
part, respectively.

For the background part one finds immediately
\begin {equation}
\label {wsbs}
w_\alpha ^{\rm B}(t_1,t_2)=w_\alpha ^{\rm B}(t_1-t_2)=N_\alpha \delta_{\alpha
\alpha }\delta(t_1-t_2)
\qquad 2\pi {\cal S}^{\rm B}(\omega )=N_\alpha \delta_{\alpha \alpha } \;,
\end {equation}
as expected for the spectrum of white noise.
Although $\delta_{\alpha \alpha }=1$, it has explicitly been kept since it
becomes divergent if the index $\alpha $ is not purely discrete, e.g.,
for plane wave modes.
For such modes, a state with ${\cal M}$, ${\cal N}$ being diagonal is an
idealization just as a plane wave coherent state.
It will be shown in the next section how this problem can be bypassed
in a simple physical way.

The `mixed' correlation function $w_\alpha ^{\rm I}(t_1,t_2)$ can be reduced to
an
expression containing only the atomic operators $\sigma ^\pm_{\textstyle t}$ by
means
of the formulae
\begin {eqnarray}
\langle \,b_\alpha ^\dagger (t_1)\sigma ^-_{\textstyle t_2}\,\rangle
&=&\vartheta (t_2-t_1)\cdot \sqrt {\gamma _\alpha }\,\langle \,
  N_\alpha [\sigma ^+_{\textstyle t_1},\sigma ^-_{\textstyle t_2}] -
  M_\alpha ^*[\sigma ^-_{\textstyle t_1},\sigma ^-_{\textstyle t_2}] \,\rangle
\\
\langle \,\sigma ^+_{\textstyle t_1}b_\alpha (t_2)\,\rangle &=&\vartheta
(t_1-t_2)\cdot \sqrt {\gamma _\alpha }\,\langle \,
  N_\alpha [\sigma ^+_{\textstyle t_1},\sigma ^-_{\textstyle t_2}] -
  M_\alpha [\sigma ^+_{\textstyle t_1},\sigma ^+_{\textstyle t_2}] \,\rangle
\end {eqnarray}
which can be derived in a similar way as in the one-dimensional case.

The remaining correlation functions of the $\sigma ^\pm$
can be evaluated in the stationary limit using the quantum regression
theorem and the atomic master equation.
Since the calculations for the one-dimensional case in \cite{gabu}
can be carried over, except for some obvious modifications,
we just give the results.
The form of the master equation
\begin {eqnarray}
\dot\rho&=&
\gamma (N+1)\Bigl(\sigma ^-\rho\sigma ^+ -
  {1\over2}
  \{\sigma ^+\sigma ^-,\rho\}\Bigr)
+\gamma N\Bigl(\sigma ^+\rho\sigma ^- -
  {1\over2}
  \{\sigma ^-\sigma ^+,\rho\}\Bigr)
\nonumber \\
&&-\gamma M\,\sigma ^+\rho\sigma ^+ -\gamma M^*\,\sigma ^-\rho\sigma ^-
\end {eqnarray}
remains unchanged, but the parameters $\gamma $, $M$, and $N$ now are defined
as
\begin {equation}
\label {gmn}
\gamma =\sum_\alpha \gamma _\alpha \qquad
M=\sum_\alpha {\gamma _\alpha \over\gamma }M_\alpha \qquad
N=\sum_\alpha {\gamma _\alpha \over\gamma }N_\alpha \;.
\end {equation}
The possible values of $M$ and $N$ are restricted by the inequality
\begin {equation}
\label {mlen1}
|M|^2\le N(N+1) \;,
\end {equation}
which follows from the relations $|M_\alpha |^2=N_\alpha (N_\alpha +1)$
[cf.\ Eq.~(\ref {emmenn})].

In the one-dimensional case $\gamma $, $M$, and $N$ defined above
agree with those appearing in Eq.~(\ref {1dim}),
and for pure states --- which we have been considering above ---
the equality sign would hold in Eq.~(\ref {mlen1}).
The full range of the parameters $M$ and $N$ can also be realized in this case
if one uses mixed states.

By absorbing a phase into the atomic states if necessary,
$M$ can be chosen real and positive.
With this convention one gets for $\tau \ge0$
(the relation $w_\alpha ^{\rm S,\,I}(\tau )=w_\alpha ^{\rm S,\,I}(-\tau )^*$
yields the corresponding
expressions for $\tau <0$)
\begin {eqnarray}
\label {wxas}
w_\alpha ^{\rm S}(\tau )&=&{\gamma _\alpha \over2}{N\over2N+1}
  \left\{{\rm e}^{-\gamma _+\tau }+{\rm e}^{-\gamma _-\tau }\right\} \\
\label {wxai}
w_\alpha ^{\rm I}(\tau )&=&-{\gamma _\alpha \over2}{1\over2N+1}
  \left\{{\rm e}^{-\gamma _+\tau }(N_\alpha +M_\alpha )+{\rm e}^{-\gamma _-\tau
}(N_\alpha -M_\alpha )\right\} \;,
\end {eqnarray}
where
\begin {equation}
\label {gpm}
\gamma _\pm=(N\pm M+{1\over2})\,\gamma \;.
\end {equation}
The complete spectrum is the Fourier transform of
\begin {eqnarray}
w_\alpha (\tau )&=&N_\alpha \delta_{\alpha \alpha }\delta(\tau ) +
 {\gamma _\alpha \over2}{1\over2N+1} \nonumber \\
&&\;\times\left\{
 {\rm e}^{-\gamma _+\tau }(N-N_\alpha -M_\alpha )+{\rm e}^{-\gamma _-\tau
}(N-N_\alpha +M_\alpha )\right\} \;,
\end {eqnarray}
\begin {eqnarray}
\label {essa}
2\pi\,{\cal S}_\alpha (\omega )&=&N_\alpha \delta_{\alpha \alpha }
+{\gamma _\alpha \over2N+1}\,\sum_\pm{\gamma _\pm\over(\omega -\omega
_0)^2+\gamma _\pm^2}
  \left(N-N_\alpha \mp\mathop {\rm Re}M_\alpha \right) \nonumber \\
&&+\;2M\gamma _\alpha \gamma ^2\,{\omega -\omega _0\over
  [(\omega -\omega _0)^2+\gamma _+^2][(\omega -\omega _0)^2+\gamma
_-^2]}\mathop {\rm Im}M_\alpha \;.
\end {eqnarray}
For the spectrum of all modes, ${\cal S}(\omega )=\sum_\alpha {\cal S}_\alpha
(\omega )$,
one obtains
\begin {equation}
2\pi\,{\cal S}(\omega )=N+{M\gamma \over2N+1}\left\{
{\gamma _-\over(\omega -\omega _0)^2+\gamma _-^2}-{\gamma _+\over(\omega
-\omega _0)^2+\gamma _+^2}\right\}
\;,
\end {equation}
an expression that coincides with the spectrum in the one-dimensional case.
This shows that, as long as only this kind of spectrum is observed,
all states satisfying (\ref {3dim}) are equivalent to mixed states of
the one-dimensional type (\ref {1dim});
c.f.\ the remark following Eq.~(\ref {mlen1}).

In contrast to ${\cal S}(\omega )$, ${\cal S}_\alpha (\omega )$ shows new
features, namely in
general the spectra are asymmetric and the sign and the relative weight
of the peaks depend on $M_\alpha $ and $N_\alpha $.
The asymmetries are caused by the phase of $M_\alpha $ (this
phase has a physical meaning as it is the relative phase of $M_\alpha $
and $M$) since the symmetric and antisymmetric parts of
${\cal S}_\alpha (\omega _0+\omega ')$ are the Fourier transforms of the real
and
imaginary part of $w_\alpha (\tau )$, respectively.

In figure~\ref{i} the spectrum (\ref {essa}) is plotted for $N_\alpha =N$ and
increasing values of $\varphi =\arg M_\alpha $.
For $\varphi =0$ one obtains the spectrum of the one-dimensional case
consisting of two Lorentzians, a positive narrow peak of width $\gamma _-$
and a negative broad peak of width $\gamma _+$.
For $0<\varphi <\pi$ the spectrum is asymmetric with respect to $\omega _0$,
the asymmetry being maximal for $\varphi =\pi/2$.
The relative weights of the symmetric contributions to the
spectrum decrease and vanish for $\varphi =\pi/2$.
For $\pi/2<\varphi \le\pi$, the weights increase again
but the signs of the two Lorentzians are interchanged.
The spectra for $\pi\le\varphi \le2\pi$ coincide with those for
$2\pi-\varphi $ mirrored at $\omega =\omega _0$.

For $N_\alpha =0$ (and consequently $M_\alpha =0$) the background and
interference parts of the spectra vanish.
As the scattered part does not depend an $N_\alpha $ and $M_\alpha $, the
spectra
take the shape of the dotted line in figure~\ref{i}
(the dashed line now representing ${\cal S}=0$).

Spectra for $N_\alpha >N$ are plotted in figure~\ref{ii}. As compared to those
with
$N_\alpha =N$, the positive peaks are attenuated while the depth of the
negative peaks increases.

\section{Example: Squeezed plane waves}
\label{III}
To be more specific, we shall now deal with plane wave modes in greater
detail.
These modes are specified by their wave-vector ${\bf k}$ and their polarization
$\lambda =1,\,2$, i.e., by $\omega =ck$ and $\alpha =({\bf \hat k},\lambda )$
with ${\bf \hat k}={\bf k}/k$.
In the dipole and rotating-wave approximation the interaction Hamiltonian
is given by
\begin {equation}
H_{\rm AF}=-{\bf d}\mathbin {\hbox {\boldmath $\cdot $}}{\bf E}{}^{(-)}\,\sigma
^--{\bf d}^*\mathbin {\hbox {\boldmath $\cdot $}}{\bf E}{}^{(+)}\,\sigma ^+ \;,
\end {equation}
where ${\bf E}{}^{(+)}={\bf E}{}^{(-)}{}^\dagger $ denotes the positive
frequency part of the
electric field at the position ${\bf r}=0$ of the atom,
\begin {equation}
{\bf E}{}^{(+)}={{\rm i}\over(2\pi)^{3\over2}}\int{\rm d}^3{\bf k}\,
\left({\hbar\omega \over2\epsilon_0}\right)^{\!1/2}
\!\!\sum_{\lambda =1,\,2}\hbox {\boldmath $\varepsilon $}^*_{{\bf \hat
k},\lambda }\,a_{{\bf k},\lambda }\;,
\end {equation}
and ${\bf d}=\langle-|\,{\bf D}\,|+\rangle$ is a matrix element of the atomic
dipole operator $\bf D$ (by parity, the static dipole moments in the states
$|\pm\rangle$ vanish).

In order to apply the formulae of the previous section, the sums over $\alpha $
and the Kronecker symbols $\delta_{\alpha _1\alpha _2}$ have to be replaced
according to
\begin {equation}
\sum_\alpha \;\longrightarrow\; \int_{S^2}{\rm d}^2{\bf \hat k}\sum_{\lambda
=1,2} \qquad
\delta_{\alpha _1\alpha _2} \;\longrightarrow\;
  \delta_{\lambda _1\lambda _2}\,\delta^2({\bf \hat k}_1,{\bf \hat k}_2) \;,
\end {equation}
where ${\rm d}^2{\bf \hat k}$ denotes the area element on the unit sphere and
$\delta^2({\bf \hat k}_1,{\bf \hat k}_2)$ the corresponding $\delta$-function.
By means of the identity
\begin {equation}
\delta^3({\bf k}_1-{\bf k}_2)=k_1^{-2}\,\delta(k_1-k_2)\,\delta^2({\bf \hat
k}_1,{\bf \hat k}_2)
\end {equation}
one can write
\begin {equation}
[a_{{\bf k}_1,\lambda _1},a^\dagger _{{\bf k}_2,\lambda _2}]=
\delta^3({\bf k}_1-{\bf k}_2)\delta_{\lambda _1\lambda _2}=
{c^3\over\omega _1^2}\cdot \delta^2({\bf \hat k}_1,{\bf \hat
k}_2)\delta_{\lambda _1\lambda _2}
\cdot \delta(\omega _1-\omega _2)
\end {equation}
and by comparing with Eq.~(\ref {acom}) one sees that operators
with commutation relations analogous to those of the $a_\alpha (\omega )$
can be defined by
\begin {equation}
a_{{\bf \hat k},\lambda }(\omega )\equiv\omega c^{-{3\over2}}\;a_{{\bf
k},\lambda }\;.
\end {equation}
The coupling coefficients appearing in a representation of $H_{\rm AF}$
in the form of Eq.~(\ref {haf}) are therefore given by
\begin {equation}
\label {cou}
g_{{\bf \hat k},\lambda }(\omega )={1\over2\pi}
\left(\omega ^3\over2\epsilon_0\hbar c^3\right)^{\!1/2}
  \hbox {\boldmath $\varepsilon $}_{{\bf \hat k},\lambda }\mathbin {\hbox
{\boldmath $\cdot $}}{\bf d}
\equiv \sqrt{\gamma _{{\bf \hat k},\lambda }(\omega )}\,{\rm e}^{{\rm
i}\phi_{{\bf \hat k},\lambda }} \;.
\end {equation}
Note that their phases do not depend on $\omega $.

As already noted, a squeezed state with ${\cal M}$, ${\cal N}$ being diagonal
for plane waves leads to a divergent expression in Eq.~(\ref {wsbs}).
One could try to circumvent this problem by subtracting the divergent
background part, but a physically more satisfactory solution is to
use an improved definition of the spectrum which avoids divergences
automatically.

A suitable quantity for modeling the spectrum actually observed in an
experiment is the spectrally resolved energy flux through a (small)
surface $\cal A$ centered at ${\bf r}={\bf r}_0$,
\begin {equation}
{\cal S}_{{\cal A},{\bf r}_0}(\omega )=\int _{({\cal A},\,{\bf r}_0)}{\bf
d}\hbox {\boldmath $\sigma $}\mathbin {\hbox {\boldmath $\cdot $}}\langle
\,{\bf S}({\bf r},\omega )\,\rangle \;,
\end {equation}
where the operator ${\bf S}({\bf r},\omega )$ represents the
`spectral Poynting vector', i.e., the spectral energy flux density.
More realistically, one could also use a direction-sensitive spectral
analyzer, e.g., an analyzer admitting only directions in a certain
cone in ${\bf k}$ space, such that only radiation with directions from
this cone is observed.
It can be shown that, as physically expected, the scattered and
interference parts of the spectrum are not influenced by this
as long as the cone contains the line between the atom and the detector.
For the background part the directional selection would simply result in a
restriction of the integration over ${\bf \hat k}$ in Eq.~(\ref {back}) below
to
this cone.

For the operator ${\bf S}({\bf r},\omega )$
we shall use the expression (cf.\ \cite{mawo})
\begin {equation}
\label {spv}
{\bf S}({\bf r},\omega )=\lim_{T\to\infty}{\epsilon_0c^2\over T}\,
\widetilde{\bf E}{}^{(-)}_T({\bf r},\omega )\mathbin {\hbox {\boldmath $\times
$}}\widetilde{\bf B}{}^{(+)}_T({\bf r},\omega )
\;+\;\hbox{h.c.} \;,
\end {equation}
where $\widetilde{\bf E}{}^{(\pm)}_T({\bf r},\omega )$
--- and analogously $\widetilde{\bf B}{}^{(\pm)}_T({\bf r},\omega )$ ---
is defined by
\begin {equation}
\widetilde{\bf E}{}^{(\pm)}_T({\bf r},\omega )={1\over\sqrt{2\pi}}\int_0^T{\rm
d}t\,
{\rm e}^{\pm{\rm i}\omega t}\,{\bf E}{}^{(\pm)}({\bf r},t) \;,
\end {equation}
${\bf E}{}^{(\pm)}({\bf r},t)$ and ${\bf B}{}^{(\pm)}({\bf r},t)$ being the
Heisenberg operators
of the positive and negative frequency parts of the (transversal) electric
and magnetic fields,
\begin {eqnarray}
{\bf E}{}^{(+)}({\bf r},t)&=&{{\rm i}\over(2\pi)^{3\over2}}\int{\rm d}^3{\bf
k}\,
  \left({\hbar\omega \over2\epsilon_0}\right)^{\!1/2}
  \!\!\sum_{\lambda =1,\,2}\hbox {\boldmath $\varepsilon $}^*_{{\bf \hat
k},\lambda }\,{\rm e}^{{\rm i}{\bf k}\mathbin {\hbox {\boldmath $\scriptstyle
\cdot $}}{\bf r}}\;
  a_{{\bf k},\lambda }(t) \\
{\bf B}{}^{(+)}({\bf r},t)&=&{{\rm i}\over c\,(2\pi)^{3\over2}}\int{\rm
d}^3{\bf k}\,
  \left({\hbar\omega \over2\epsilon_0}\right)^{\!1/2}
  \!\!\sum_{\lambda =1,\,2}{\bf \hat k}\mathbin {\hbox {\boldmath $\times
$}}\hbox {\boldmath $\varepsilon $}^*_{{\bf \hat k},\lambda }\,{\rm e}^{{\rm
i}{\bf k}\mathbin {\hbox {\boldmath $\scriptstyle \cdot $}}{\bf r}}\;
  a_{{\bf k},\lambda }(t) \;.
\end {eqnarray}
Introducing the correlation function%
\ \cite{minus}
\begin {eqnarray}
w_{{\cal A},{\bf r}_0}(t_1,t_2)=\epsilon_0c^2\int _{({\cal A},\,{\bf r}_0)}{\bf
d}\hbox {\boldmath $\sigma $}\mathbin {\hbox {\boldmath $\cdot $}}
  {\rm e}^{-{\rm i}\omega _0(t_1-t_2)}\,\langle \,
  {\bf E}{}^{(-)}({\bf r},t_1)\mathbin {\hbox {\boldmath $\times $}}{\bf
B}{}^{(+)}({\bf r},t_2) \hphantom{-\,\rangle }&&\nonumber \\
-{\bf B}{}^{(-)}({\bf r},t_1)\mathbin {\hbox {\boldmath $\times $}}{\bf
E}{}^{(+)}({\bf r},t_2)\,\rangle &&
\end {eqnarray}
and its stationary limit $w_{{\cal A},{\bf r}_0}(\tau
)=\lim_{t_2\to\infty}w_{{\cal A},{\bf r}_0}(\tau +t_2,t_2)$,
${\cal S}_{{\cal A},{\bf r}_0}(\omega )$ can be written as a Fourier transform
in the same way as
${\cal S}_\alpha (\omega )$ in Eqs.~(\ref {sft12}) and (\ref {sft}),
\begin {eqnarray}
{\cal S}_{{\cal A},{\bf r}_0}(\omega )&=&\lim_{T\to\infty}{1\over2\pi T}
  \int_0^T{\rm d}t_1\int_0^T{\rm d}t_2\,{\rm e}^{-{\rm i}(\omega -\omega
_0)(t_1-t_2)}
  \,w_{{\cal A},{\bf r}_0}(t_1,t_2) \nonumber \\
&=&{1\over\pi}\mathop {\rm Re}\int_0^\infty{\rm d}\tau \,{\rm e}^{-{\rm
i}(\omega -\omega _0)\tau }\,w_{{\cal A},{\bf r}_0}(\tau ) \;.
\end {eqnarray}

Since the time evolution operator
$U(t)=\exp\left(-i\hbar^{-1}(H_{\rm A}+H_{\rm F})t\right)\*U_{\rm I}(t,0)$
transforms $b_\alpha (s)$ into
\begin {equation}
U(t)^\dagger \,b_\alpha (s)\,U(t)={\rm e}^{-{\rm i}\omega _0t}\,b_\alpha
(s+t)_{\textstyle t} \;,
\end {equation}
it follows from Eqs.~(\ref {bsa}) and (\ref {cou}) that the $a_{{\bf k},\lambda
}(t)$ are
related to the $b_{{\bf \hat k},\lambda }(t+s)_{\textstyle t}$ by
\begin {equation}
\label {arelb}
{\rm e}^{{\rm i}\phi_{{\bf \hat k},\lambda }}\,{\rm e}^{-{\rm i}\omega
_0(t+s)}\,b_{{\bf \hat k},\lambda }(t+s)_{\textstyle t}=
{1\over\sqrt{2\pi}}\int_0^\infty{\rm d}\omega
\left(\omega \over\omega _0\right)^p
\omega c^{-{3\over2}}\,{\rm e}^{-{\rm i}\omega s}\,a_{{\bf k},\lambda }(t)
\end {equation}
with $p=3/2$.
Within the scope of the Markov approximation, this equation remains valid
also for $p\ne3/2$ since the factor $\omega /\omega _0$ can be replaced by
unity just as $\gamma _\alpha (\omega )/\gamma _\alpha (\omega _0)$ in
Eq.~(\ref {bcom})
(cf.\ \cite{gabu}, Ch.~8.1).

Because of the linearity of the above relation, the decomposition of the
$b_{{\bf \hat k},\lambda }(s)_{\textstyle t}$ according to Eq.~(\ref {bsat}),
\begin {eqnarray}
b_{{\bf \hat k},\lambda }(t+s)_{\textstyle t}&=&b_{{\bf \hat k},\lambda
}(t+s)+\vartheta (-s)\,\sqrt {\gamma _{{\bf \hat k},\lambda }}\,\sigma
^-_{\textstyle t+s} \nonumber \\
&\equiv&b_{{\bf \hat k},\lambda }(t+s)_{\textstyle t}{}^{(1)}+b_{{\bf \hat
k},\lambda }(t+s)_{\textstyle t}{}^{(2)} \;,
\end {eqnarray}
leads to an equivalent decomposition of the $a_{{\bf k},\lambda }(t)$ as well
as of the
fields ${\bf E}{}^{(+)}({\bf r},t)$ and ${\bf B}{}^{(+)}({\bf r},t)$.
Thus the correlation function $w_{{\cal A},{\bf r}_0}(t_1,t_2)$ can be written
as a sum $w_{{\cal A},{\bf r}_0}=w_{{\cal A},{\bf r}_0}^{\rm B}+w_{{\cal
A},{\bf r}_0}^{\rm I}+w_{{\cal A},{\bf r}_0}^{\rm S}$ analogous to Eq.~(\ref
{bis})
for $w_\alpha $.
In the following, we shall denote by $\langle \,\ldots\,\rangle ^{\rm
B,\,I,\,S}$
the sum of all the terms in $\langle \,\ldots\,\rangle $ that contribute
to $w_{{\cal A},{\bf r}_0}^{\rm B,\,I,\,S}$, i.e., we set
\begin {displaymath}
\langle \,b_1^\dagger b_2\,\rangle ^{\rm B}=\langle \,b_1^\dagger
{}^{(1)}b_2{}^{(1)}\,\rangle \qquad
\langle \,b_1^\dagger b_2\,\rangle ^{\rm S}=\langle \,b_1^\dagger
{}^{(2)}b_2{}^{(2)}\,\rangle \;,
\end {displaymath}\begin {equation}
\langle \,b_1^\dagger b_2\,\rangle ^{\rm I}=
  \langle \,b_1^\dagger {}^{(1)}b_2{}^{(2)}+b_1^\dagger
{}^{(2)}b_2{}^{(1)}\,\rangle \;,
\end {equation}
with $b_i$ standing for $b_{{\bf \hat k}_i,\lambda _i}(t_i+s_i)_{\textstyle
t_i}$.

Using Eq.~(\ref {arelb}) the correlation function of the background part
can easily be calculated.
With
\begin {equation}
2\,{\bf \Xi }({\bf \hat k}_1,\lambda _1;{\bf \hat k}_2,\lambda _2)\equiv
\hbox {\boldmath $\varepsilon $}_{{\bf \hat k}_1,\lambda _1}\mathbin {\hbox
{\boldmath $\times $}}({\bf \hat k}_2\mathbin {\hbox {\boldmath $\times
$}}\hbox {\boldmath $\varepsilon $}^*_{{\bf \hat k}_2,\lambda _2})-
({\bf \hat k}_1\mathbin {\hbox {\boldmath $\times $}}\hbox {\boldmath
$\varepsilon $}_{{\bf \hat k}_1,\lambda _1})\mathbin {\hbox {\boldmath $\times
$}}\hbox {\boldmath $\varepsilon $}^*_{{\bf \hat k}_2,\lambda _2}
\end {equation}
one finds
\begin {eqnarray}
\label {back}
w_{{\cal A},{\bf r}_0}^{\rm B}(t_1-t_2)&=&
  {\hbar\omega _0\over\lambda _0^2}\int _{({\cal A},\,{\bf r}_0)}{\bf d}\hbox
{\boldmath $\sigma $}\mathbin {\hbox {\boldmath $\cdot $}}
  \int{\rm d}^2{\bf \hat k}_1\int{\rm d}^2{\bf \hat k}_2\sum_{\lambda
_1,\,\lambda _2}\;
  {\bf \Xi }({\bf \hat k}_1,\lambda _1;{\bf \hat k}_2,\lambda _2) \nonumber \\
&&\quad\times\;\exp\left(-{\rm i}\phi_{{\bf \hat k}_1,\lambda _1}+
    {\rm i}\phi_{{\bf \hat k}_2,\lambda _2}\right)\,
  \exp\left({\rm i}\omega _0c^{-1}({\bf \hat k}_1-{\bf \hat k}_2)\mathbin
{\hbox {\boldmath $\cdot $}}{\bf r}\right) \nonumber \\
&&\quad\times\;\langle \,b_{{\bf \hat k}_1,\lambda _1}^\dagger (t_1-c^{-1}{\bf
\hat k}_1\mathbin {\hbox {\boldmath $\cdot $}}{\bf r})_{\textstyle t_1}\,
  b_{{\bf \hat k}_2,\lambda _2}(t_2-c^{-1}{\bf \hat k}_2\mathbin {\hbox
{\boldmath $\cdot $}}{\bf r})_{\textstyle t_2}\,\rangle ^{\rm B} \nonumber \\
&=&\delta(t_1-t_2)\cdot {\hbar\omega _0\over\lambda _0^2}\int _{({\cal
A},\,{\bf r}_0)}{\bf d}\hbox {\boldmath $\sigma $}\mathbin {\hbox {\boldmath
$\cdot $}}
  \int{\rm d}^2{\bf \hat k}\sum_\lambda N_{{\bf \hat k},\lambda }\,{\bf \hat
k}\;,
\end {eqnarray}
where $\lambda _0=2\pi c/\omega _0$ is the wavelength corresponding to the
atomic
transition frequency.

The calculation of the interference and scattered part can be performed
with the aid of the asymptotic expansion \cite{bowo,ueb}
\begin {equation}
\label {asymp}
\int{\rm d}^2{\bf \hat k}\,{\rm e}^{-{\rm i}{\bf k}\mathbin {\hbox {\boldmath
$\scriptstyle \cdot $}}{\bf r}}\,f({\bf \hat k})
\;\buildrel r\to\infty\over\sim\;
{2\pi{\rm i}\over kr}
  \left\{{\rm e}^{-{\rm i}kr}\,f({\bf \hat r})-{\rm e}^{{\rm i}kr}\,f(-{\bf
\hat r})\right\} +
  O\left(r^{-2}\right)
\end {equation}
being valid as long as $f({\bf \hat k})$ is sufficiently smooth.
Note that this condition is violated for the background part%
\ \cite{careful}
calculated above
where $f({\bf \hat k})\propto\delta^2({\bf \hat k},{\bf \hat k}')$.
By first applying (\ref {asymp}) to either integral over ${\bf \hat k}$ in
$w_{{\cal A},{\bf r}_0}^{\rm I,\,S}$,
keeping only terms of the leading order $r^{-1}$, and then using
Eq.~(\ref {arelb}), one obtains
\begin {eqnarray}
w_{{\cal A},{\bf r}_0}^{\rm I,\,S}(t_1,t_2)&=&
  {\hbar\omega _0\over r^2}\int _{({\cal A},\,{\bf r}_0)}{\bf d}\hbox
{\boldmath $\sigma $}\mathbin {\hbox {\boldmath $\cdot $}}
  \sum_{\xi_1=\pm1}\sum_{\xi_2=\pm1}\sum_{\lambda _1,\,\lambda _2}\xi_1\xi_2\;
  {\bf \Xi }(\xi_1{\bf \hat r},\lambda _1;\xi_2{\bf \hat r},\lambda _2)
\nonumber \\
&&\quad\times\;\exp\left(-{\rm i}\phi_{\xi_1{\bf \hat r},\lambda _1}+
    {\rm i}\phi_{\xi_2{\bf \hat r},\lambda _2}\right)\,
  \exp\left({\rm i}\omega _0c^{-1}(\xi_1-\xi_2)r\right) \nonumber \\
&&\quad\times\langle \,b_{\xi_1{\bf \hat r},\lambda _1}^\dagger
(t_1-c^{-1}\xi_1r)_{\textstyle t_1}\,
  b_{\xi_2{\bf \hat r},\lambda _2}(t_2-c^{-1}\xi_2r)_{\textstyle t_2}\,\rangle
^{\rm I,\,S} \;.
\end {eqnarray}
Due to
\begin {equation}
{\bf \Xi }({\bf \hat k},\lambda _1;-{\bf \hat k},\lambda _2)=0 \qquad
{\bf \Xi }({\bf \hat k},\lambda _1;{\bf \hat k},\lambda _2)={\bf \hat
k}\,\delta_{\lambda _1\lambda _2}
\end {equation}
the sums are actually running only over $\xi_1=\xi_2$ and $\lambda _1=\lambda
_2$.
Further, since $\langle \,\ldots\,\rangle ^{\rm I,\,S}$ contains at least one
factor
$\vartheta (t_i-t_i+c^{-1}\xi_ir)=\vartheta (\xi_i)$, only the term with
$\xi_1=\xi_2=+1$
survives.
For a surface which is sufficiently flat and whose diameter is small
compared to its distance to the atom,
${\bf r}$ can be approximated by ${\bf r}_0$, the integration over $({\cal
A},{\bf r}_0)$
resulting in a multiplication with the oriented area $\hbox {\boldmath $\cal
A$}=\int _{({\cal A},\,{\bf r}_0)}{\bf d}\hbox {\boldmath $\sigma $}$.
So we finally have, with $w_{{\bf \hat r},\lambda }^{\rm I,\,S}$ defined as in
Eqs.~(\ref {wxas}) and
(\ref {wxai}),
\begin {equation}
w_{{\cal A},{\bf r}_0}^{\rm I,\,S}(t_1,t_2)=
\hbar\omega _0\,{\hbox {\boldmath $\cal A$}\mathbin {\hbox {\boldmath $\cdot
$}}{\bf \hat r}_0\over r_0^2}\sum_\lambda
w_{{\bf \hat r}_0,\lambda }^{\rm I,\,S}(t_1-c^{-1}r_0,\,t_2-c^{-1}r_0)
\end {equation}
and
\begin {equation}
\label {wxfis}
w_{{\cal A},{\bf r}_0}^{\rm I,\,S}(t)=
\hbar\omega _0\,{\hbox {\boldmath $\cal A$}\mathbin {\hbox {\boldmath $\cdot
$}}{\bf \hat r}_0\over r_0^2}\sum_\lambda
w_{{\bf \hat r}_0,\lambda }^{\rm I,\,S}(t) \;.
\end {equation}
By inserting Eq.~(\ref {cou}) one explicitly finds for $\gamma $, $M$, and $N$
in Eq.~(\ref {gmn})
\begin {equation}
\gamma ={\omega _0^3\,|{\bf d}|^2\over3\pi\epsilon_0\hbar c^3} \;,
\end {equation}
(this is the Einstein $A$ coefficient for a dipole transition)
\begin {equation}
\label {mn}
M={3\over8\pi}\int\!{\rm d}^2{\bf \hat k}\sum_\lambda |\hbox {\boldmath
$\varepsilon $}_{{\bf \hat k},\lambda }\mathbin {\hbox {\boldmath $\cdot
$}}{\bf \hat d}|^2
  M_{{\bf \hat k},\lambda } \qquad
N={3\over8\pi}\int\!{\rm d}^2{\bf \hat k}\sum_\lambda |\hbox {\boldmath
$\varepsilon $}_{{\bf \hat k},\lambda }\mathbin {\hbox {\boldmath $\cdot
$}}{\bf \hat d}|^2
  N_{{\bf \hat k},\lambda }
\end {equation}
and the three parts of the spectrum read
\begin {eqnarray}
\label {sxfb}
&&{2\pi\over\hbar\omega _0}\,{\cal S}_{{\cal A},{\bf r}_0}^{\rm B}(\omega )=
  \lambda _0^{-2}\int _{({\cal A},\,{\bf r}_0)}{\bf d}\hbox {\boldmath $\sigma
$}\mathbin {\hbox {\boldmath $\cdot $}}\int\!{\rm d}^2{\bf \hat k}\sum_\lambda
N_{{\bf \hat k},\lambda }{\bf \hat k}\\
\label {sxfs}
&&{2\pi\over\hbar\omega _0}\,{\cal S}_{{\cal A},{\bf r}_0}^{\rm S}(\omega )=
  {3\,\hbox {\boldmath $\cal A$}\mathbin {\hbox {\boldmath $\cdot $}}{\bf \hat
r}_0\over8\pi r_0^2}\,{N\gamma \over2N+1}\,
  \left(1-|{\bf \hat d}\mathbin {\hbox {\boldmath $\cdot $}}{\bf \hat
r}_0|^2\right)\,
  \sum_\pm{\gamma _\pm\over(\omega -\omega _0)^2+\gamma _\pm^2} \\
\label {sxfi}
&&{2\pi\over\hbar\omega _0}\,{\cal S}_{{\cal A},{\bf r}_0}^{\rm I}(\omega )=
  {3\,\hbox {\boldmath $\cal A$}\mathbin {\hbox {\boldmath $\cdot $}}{\bf \hat
r}_0\over8\pi r_0^2}\,
  \sum_\lambda \left|\hbox {\boldmath $\varepsilon $}_{{\bf \hat r}_0,\lambda
}\mathbin {\hbox {\boldmath $\cdot $}}{\bf \hat d}\right|^2 \nonumber \\
&&\qquad\qquad\times\Biggl\{
  {\gamma \over2N+1}\sum_\pm{\gamma _\pm\over(\omega -\omega _0)^2+\gamma
_\pm^2}
  \left(-N_{{\bf \hat r}_0,\lambda }\mp\mathop {\rm Re}M_{{\bf \hat
r}_0,\lambda }\right) \nonumber \\
&&\qquad\qquad\hphantom{\times\Biggl\{}
  +\;2M\gamma ^3{\omega -\omega _0\over[(\omega -\omega _0)^2+\gamma
_+^2][(\omega -\omega _0)^2+\gamma _-^2]}
  \mathop {\rm Im}M_{{\bf \hat r}_0,\lambda }\Biggr\} \;,
\end {eqnarray}
with $\lambda _0$ the wavelength of the atomic transition frequency.
We note the direction dependence of the spectrum and the $1/r^2$
dependence of ${\cal S}_{{\cal A},{\bf r}_0}^{\rm S}$ and ${\cal S}_{{\cal
A},{\bf r}_0}^{\rm I}$.

Eq.~(\ref {wxfis}) shows that, as far as the interference
and scattered parts are concerned, the spectrum in
Eqs.~(\ref {sxfb}) -- (\ref {sxfi})
is a sum of spectra for fixed polarization.
The latter correspond to a spectrum for fixed $\alpha $ as in Section~\ref{II}
and can be obtained by omitting the sum over $\lambda $ in Eq.~(\ref {sxfi})
and
replacing the factor $1-|{\bf \hat d}\mathbin {\hbox {\boldmath $\cdot $}}{\bf
\hat r}_0|^2$ by $|\hbox {\boldmath $\varepsilon $}_{{\bf \hat r}_0,\lambda
}\mathbin {\hbox {\boldmath $\cdot $}}{\bf \hat d}|$
in Eq.~(\ref {sxfs}).

The basic features of the spectra are the same as in the general case of
Section~\ref{II}.
This will now be discussed in more detail.

\section{Discussion}
\label{IV}
The spectra in Eq.~(\ref {essa}), as well as those in Eqs.~(\ref {sxfb}) --
(\ref {sxfi}),
are a sum of a background term ${\cal S}^{\rm B}$,
a scattered part ${\cal S}^{\rm S}$ and a term ${\cal S}^{\rm I}$, which can be
identified as an interference part.
That this term is indeed due to interference (note that all radiation
is quantized) can be seen in various ways.
Formally, this is already suggested by Eqs.~(\ref {out}) and (\ref {bis}).
The last term in Eq.~(\ref {out}) is due to the presence of the atom since it
vanishes for $\gamma _\alpha =0$ and Eq.~(\ref {bis}) is due to the cross terms
of this
with $b_\alpha $%
\cite{cavinter}.

The interference nature of ${\cal S}^{\rm I}$ becomes yet more transparent if
one
uses the spectral Poynting vector for the calculation of the spectrum, as in
Section~\ref{III}.
As pointed out there, the results for ${\cal S}^{\rm S}$
and ${\cal S}^{\rm I}$ remain the same for a direction-sensitive spectral
analyzer,
as long as the analyzer points in the direction of the atom.
If the direction-sensitive analyzer does not point in this direction one
shows by the same
arguments as in Section~\ref{III} that ${\cal S}^{\rm S}$ and
${\cal S}^{\rm I}$ become zero. This is physically expected since it means
that the corresponding light originates at the site of the atom.
Furthermore, if the vacuum is squeezed only for directions in some cone
$\cal C$, i.e., if $N_{{\bf \hat k},\lambda }$ and $M_{{\bf \hat k},\lambda }$
vanish for ${\bf \hat k}$ not
in $\cal C$, then ${\cal S}^{\rm I}$ vanishes if the
direction ${\bf \hat r}_0$ from atom to analyzer is not in the squeezing cone
$\cal C$, since then $N_{{\bf \hat r}_0,\lambda }$ and $M_{{\bf \hat
r}_0,\lambda }$ in Eq.~(\ref {sxfi}) vanish.
Moreover, ${\cal S}^{\rm I}$ depends on $N_{{\bf \hat k},\lambda }$ and
$M_{{\bf \hat k},\lambda }$
only for ${\bf \hat k}={\bf \hat r}_0$ [except for the general dependence of
$N$ and $M$
in Eq.~(\ref {mn})], and this can be interpreted as the fact that scattered
light interferes only with incident light traveling in the same direction,
just as for classical light scattering.
This can be traced back to Eq.~(\ref {asymp}).

The spectra in Eqs.~(\ref {sxfb}) -- (\ref {sxfi})
depend on the direction ${\bf \hat r}_0$ of the analyzer.
As seen from Eq.~(\ref {sxfs}), outside the squeezing cone the spectrum
consists of a narrow Lorentzian sitting on top of a broad Lorentzian.
This is a special case of the dependence on $\alpha $ in Eq.~(\ref {essa}).
As remarked earlier there is no interference outside the squeezing cone.
The analogous fact is true in the general case, as seen in Eq.~(\ref {wxai})
for ${\cal S}_\alpha ^{\rm I}$ which vanishes when the channel $\alpha $ is not
squeezed.

For multi-channel squeezing one has more parameters available than in
the single-channel case.
In addition to $N$ and $M$ of the single-channel case one now also has
$N_\alpha $ and $M_\alpha $ (or $N_{{\bf \hat k},\lambda }$ and $M_{{\bf \hat
k},\lambda }$ in the plane-wave
model), with $|M_\alpha |^2=N_\alpha (N_\alpha +1)$.
The range of $M$ and $N$ is restricted by the inequality (\ref {mlen1}),
\begin {equation}
\label {mlen}
|M|^2\le N(N+1) \,,
\end {equation}
where the equality sign holds, as easily shown, if and only if
$N_\alpha \equiv N$ and $M_\alpha \equiv M$ for all $\alpha $ with $\gamma
_\alpha \ne0$.
Only in this case does the spectrum ${\cal S}_\alpha (\omega )$ in Eq.~(\ref
{essa})
contain a peak which becomes increasingly narrow for increasing $N$
as seen from the definition of $\gamma _-$ in Eq.~(\ref {gpm})
which for constant ratio $m\equiv M[N(N+1)]^{-1/2}$ can be written as
\begin {displaymath}
\gamma _-=\left[(N+{\textstyle{1\over2}})(1-m)+{m\over8N}+
  O\left(N^{-2}\right)\right]\gamma \;.
\end {displaymath}
For $m=1$ this not only recovers the one-dimensional like case of
Gardiner \cite{gasq} with its interesting narrow peak%
\ \cite{broadinter},
but also shows that a subnatural linewidth and the above new features are
mutually exclusive.
A similar situation is found in Ref.~\cite {fismysw} for single-channel
squeezing with an additional laser.

\begin{figure}
\epsfxsize=\hsize \epsfbox {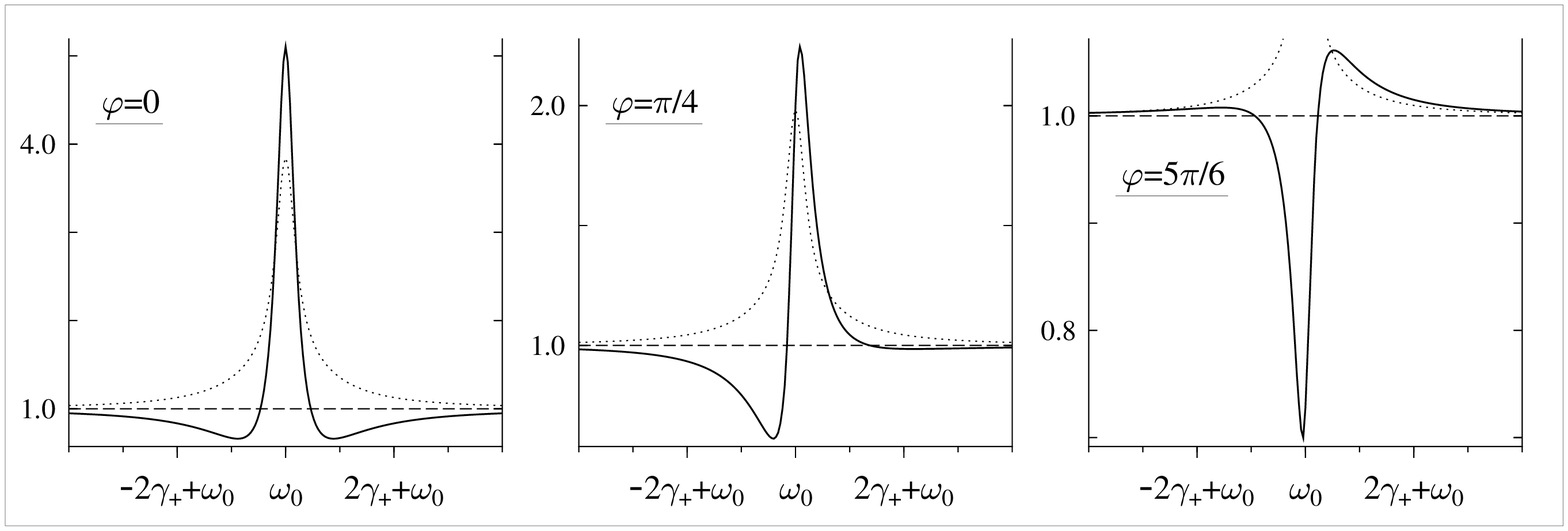}
\caption
{Typical spectra ${\cal S}_\alpha (\omega )$ for $N_\alpha =N=0.25$,
 $M^2=0.75\,N(N+1)$ and different
 phases $\varphi$ of $M_\alpha $ (solid line).
 The dashed line denotes the background part of the spectrum,
 the dotted line the sum of the background and scattered part.
 The spectra belong to the largest values of $\gamma _\alpha /\gamma $
 compatible with Eq.~(\ref {gmn}).
 The $y$ axes are in units of $N_\alpha /2\pi$.}
\label{i}
\end{figure}

\begin{figure}
\epsfxsize=\hsize \epsfbox {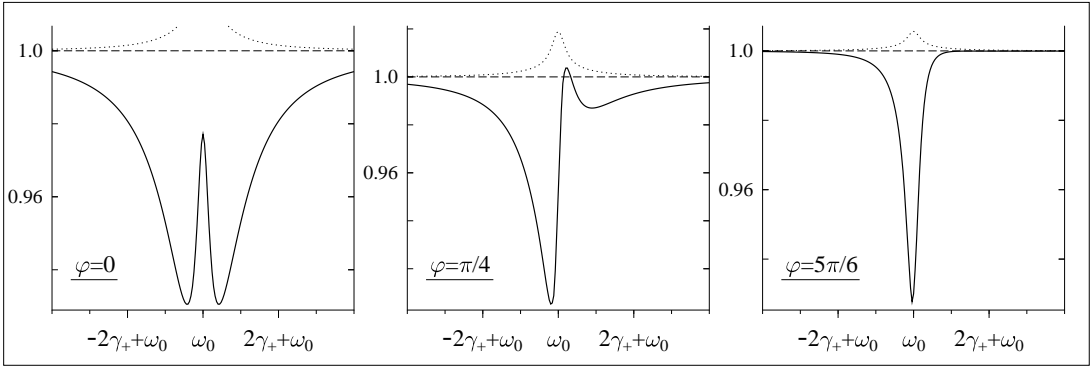}
\caption
{The spectra of figure~\protect\ref{i}, but with $N_\alpha =8N$, $N=0.25$,
 and $M^2=0.75\,N(N+1)$.}
\label{ii}
\end{figure}

If the equality does not hold in Eq.~(\ref {mlen}), then $\gamma _-$ increases
with
increasing $N$ and the corresponding peak cannot become arbitrarily narrow.
But as long as $M^2/[N(N+1)]$ is not to small
(the spectra in figure~\ref{i} and \ref{ii} belong to $M^2/[N(N+1)]=0.75$),
there are other interesting features in this case.
First of all, if $M_\alpha $ is real then the spectrum is symmetric.
However, if $M_\alpha $ is complex (since we have chosen $M$ as positive this
actually means $M_\alpha /M$ complex) then $\mathop {\rm Im}M_\alpha \ne0$ and
the spectrum is
asymmetric.
But even if $M_\alpha $ is real, new phenomena occur,
as seen in figure~\ref{i} and \ref{ii} for $\varphi =0$.
In this case the negative contribution can substantially exceed the positive
one in absolute value, as shown in figure~\ref{ii} ($\varphi =0$), and for
negative
$M_\alpha /M$ the central maximum can be absent completely.
The same is true in Eqs.~(\ref {sxfs}) and (\ref {sxfi}) for the plane wave
model.

If $M_\alpha /M$ is complex then the last term in Eq.~(\ref {essa}), which
comes
from the interference part, makes the spectrum asymmetric.
For the plane-wave model this is seen in Eq.~(\ref {sxfi}).
This asymmetry is a new phenomenon which does not occur in the
one-dimensional case.
Various typical spectra are shown in figure~\ref{i} and \ref{ii}.

\begin{figure}
\epsfxsize=\hsize \epsfbox {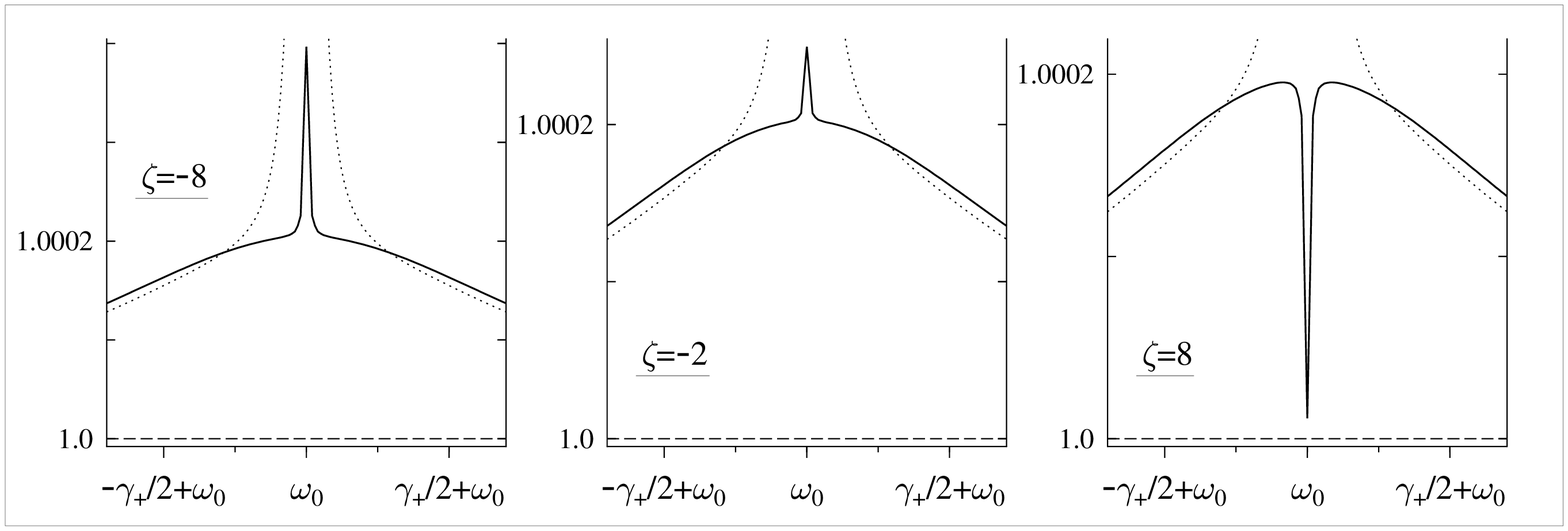}
\caption
{Spectra with $N_\alpha =(1+\zeta/1000){N^2\over2N+1}$, $N=5$,
 $M^2=0.98\,N(N+1)$ and $\arg M_\alpha =\pi$.}
\label{iii}
\end{figure}

It seems that the relative phases of $M$ and $M_\alpha $ play a similar role
for the spectra as the introduction of an additional laser with a relative
phase in the one-channel case, which has been investigated in
Refs.~\cite {smsw,swain}.
The spectra obtained in Ref.~\cite {swain} resemble those in our figures.
The similarity is particularly striking for figure~\ref{iii}%
\ \cite{eins}.
Spectra with a `pimple' and central hole burning occur also in
Ref.~\cite {swain}, and we have a similar sensitivity of this effect
on the parameters (in figure~\ref{iii} the `pimple' disappears for $\zeta=0$,
corresponding to the cancellation of the contributions from the scattered
and interference part, respectively).
It seems to us, however, that this is more a formal mathematical similarity
of two physically distinct situations since in both cases one has sums and
differences of Lorentzians and the possibility to adjust various parameters.
Physically, the spectra calculated in Ref.~\cite {swain} belong solely to the
scattered light from the atom, which is driven by the combined field of the
laser and the squeezed vacuum, and therefore these spectra
correspond to detection directions away from the driving fields.
In our case, however, these spectra result from the
interference of the radiation emitted by the atom with the squeezed
light traveling away from the atom, and in these directions the
combined light is spectrally analyzed.

In summary, we have shown that in the case of an atom in a more general
squeezed vacuum the spectrum can show new phenomena compared to the
one-dimensional like case, among them absence of the positive peak and
asymmetry.
In the plane-wave model the shape of the spectrum can become direction
dependent which, in the general case, is translated into $\alpha $ dependence.
This dependence is due to an interference effect of the squeezed light with
the scattered light.

\newpage

\begin{thebibliography}{**}
\bibitem {gasq} C.W. Gardiner,
  Phys.\ Rev.\ Lett.\ {\bf 56}, 1917 (1985).
\bibitem {gaco} C.W. Gardiner, A.S. Parkins, and M.J. Collett,
  J. Opt.\ Soc.\ Am.\ B {\bf 4}, 1863 (1987).
\bibitem {park} A.S. Parkins, in
  {\it Modern Nonlinear Optics}, edited by M. Evans and S. Kielich
  (Wiley, New York, 1993), p. 607.
\bibitem {p} P. Zhou and S. Swain,
  Opt.\ Commun.\ {\bf 131}, 153 (1996)
  J. Opt.\ Soc.\ Am.\ B {\bf 13}, 768 (1996);
  P.R. Rice and C.A. Baird,
  Phys.\ Rev.\ A {\bf 53}, 3633 (1996);
  C. Cabrillo, W.S. Smyth, S. Swain, and P. Zhou,
  Opt.\ Commun.\ {\bf 114}, 344 (1995);
  W.S. Smyth and S. Swain,
  Opt.\ Comun.\ {\bf 112}, 91 (1995);
  A. Banerjee,
  Phys.\ Rev.\ A {\bf 52}, 2472 (1995);
  R.R. Tucci,
  Opt.\ Commun.\ {\bf 118}, 241 (1995);
  Z. Ficek and B.C. Sanders,
  J. Phys.\ B {\bf 27}, 809 (1994);
  Z. Ficek, W.S. Smyth, and S. Swain,
  Opt.\ Commun.\ {\bf 110}, 555 (1994);
  S. Smart and S. Swain,
  J. Mod.\ Opt.\ {\bf 41}, 1055 (1994);
  {\bf 40}, 1939 (1993);
  Opt.\ Commun.\ {\bf 99}, 369 (1993);
  N.H. Moin and M.R.B. Wahiddin,
  Opt.\ Commun.\ {\bf 100}, 105 (1993).
\bibitem {fismysw} Z. Ficek, W.S. Smyth, and S. Swain,
  Phys.\ Rev.\ A {\bf 52}, 4126 (1995).
\bibitem {gabu} C.W. Gardiner,
  {\it Quantum Noise} (Springer, Berlin, 1992),
  Ch.~9 and 10.
\bibitem {clw} H.J. Carmichael, A.S. Lane, and D.F. Walls,
  Phys.\ Rev.\ Lett.\ {\bf 58}, 2539 (1987).
\bibitem {smsw} S. Smart and S. Swain,
  Phys.\ Rev.\ A {\bf 48}, R50 (1993).
\bibitem {swain} S. Swain
  Phys.\ Rev.\ Lett.\ {\bf 73}, 1493 (1994).
\bibitem {fd} Z. Ficek and P.D. Drummond,
  Phys.\ Rev.\ A {\bf 43}, 6247 (1991); {\bf 43}, 6258 (1991).
\bibitem {buzek} V. Bu\v zek, P.L. Knight, and K. Kudryavtsev,
  Phys.\ Rev.\ A {\bf 44}, 1931 (1991).
\bibitem {ga} C.W. Gardiner and M.J. Collett,
  Phys.\ Rev.\ A {\bf 31}, 3761 (1985);
  C.W. Gardiner, A.S. Parkins, and P. Zoller,
  {\it ibid.}\ {\bf 46}, 4363 (1992).
\bibitem {indep}
 To achieve this kind of squeezing it is not necessary for the parametric
 oscillators to be completely isolated. On the contrary, one has to
 derive the pumps of the oscillators from the same source to provide
 for the time stability of the relative phases $\arg(M_\alpha /M_\beta)$
 whose importance will emerge below.
\bibitem {spn} R. Simon, N. Mukunda, and B. Dutta,
  Phys.\ Rev.\ A {\bf 49}, 1567 (1994).
\bibitem {spnpre} R. Simon, N. Mukunda, and B. Dutta,
  {\it The Real Symplectic Groups in Quantum Mechanics and Optics}
  (Preprint quant-ph/9509002).
\bibitem {mawo} L. Mandel and E. Wolf
  {\it Optical Coherence and Quantum Optics}
  (Cambridge University Press, Cambride, 1995), Ch.~6.6.
\bibitem {minus}
  The minus sign is a consequence of the vector product.
  As $w_\alpha $ so does $w_{{\cal A},{\bf r}_0}$ satisfy the relation
  $w_{{\cal A},{\bf r}_0}(t_1,t_2)^*=w_{{\cal A},{\bf r}_0}(t_2,t_1)$.
\bibitem {bowo} M. Born and E. Wolf,
  {\it Principles of Optics} (Pergamon, Oxford, 1975).
\bibitem {ueb} A. Messiah,
  {\it Quantum Mechanics} (North-Holland, Amsterdam, 1961),
  Exercise 1 in Ch.~19.
\bibitem {careful}
  Even for the scattered and interference part the application of (\ref
{asymp})
  requires a careful analysis since in these cases  one deals with an
  $f({\bf \hat k})$ containing step functions.
  Due to this we have derived our results also by direct evaluation,
  using integration by parts and the Riemann--Lebesgue lemma.
\bibitem {cavinter}
  In connection with cavities interference is discussed in Ref.~\cite {park}.
\bibitem {broadinter}
  The broad negative peak found there comes from the interference term.
\bibitem {eins}
  Note that for the spectra in figure~\ref{iii} we have chosen a higher value
  of $M^2/[N(N+1)]$ than for those in figure~\ref{i} and \ref{ii} because
  the properties shown in this figure are clearly identifiable only for
  $\gamma _-\ll\gamma _+$ which requires $M^2/[N(N+1)]$ being close to unity.
\end{thebibliography}
\end{document}